\documentclass[aps,prd,reprint,nofootinbib]{revtex4-1}
\usepackage{amsmath,amssymb}
\usepackage[T1]{fontenc}
\usepackage[margin=1in]{geometry}
\usepackage{graphicx}
\usepackage{float}
\usepackage{microtype}
\usepackage{xcolor}

\begin{document}

\title{Analytic Quasinormal Spectrum of Effective de Sitter Space in Generalized Proca Theory}
\author{Zainab Malik}
\email{zainabmalik8115@outlook.com}
\affiliation{Institute of Applied Sciences and Intelligent Systems, H-15 Islamabad, Pakistan}

\begin{abstract}
Quasinormal modes describe the relaxation of perturbed black holes and relate ringdown observables to the background geometry. In this work we study the problem in a de Sitter setting within a generalized Proca branch that generates an effective positive cosmological constant and admits an exact de Sitter vacuum. Using this vacuum, we derive closed expressions for scalar mode frequencies and identify the change in damping behavior between light and heavy fields. The resulting formulas show explicitly how the theory parameters determine the de Sitter-like part of the spectrum.
\end{abstract}

\maketitle

\textbf{Introduction.} Quasinormal modes (QNMs) play a central role in black-hole physics. They determine the late-time ringdown after perturbations, encode linear stability information, and provide a standard tool for black-hole spectroscopy. In gravitational-wave phenomenology, QNMs are important because their frequencies and damping times carry direct information about the underlying geometry. Comprehensive reviews of the subject and its applications to black-hole spectroscopy, classical stability, and tests of gravity can be found in~\cite{Kokkotas:1999bd,Berti:2009kk,Konoplya:2011qq,Bolokhov:2025uxz}.

A key technical point is that closed-form QNM spectra are exceptional: for generic black-hole backgrounds the perturbation equations do not reduce to solvable special-function problems, and frequencies must be computed numerically (e.g., by continued fractions, direct integration, or spectral/discretization methods). Analytically tractable cases are therefore valuable benchmarks, because they isolate parameter dependences explicitly and provide high-precision reference data for broader numerical studies.

For asymptotically de Sitter black holes (BHs), the problem is more involved than in asymptotically flat spacetimes. The positive cosmological constant introduces a cosmological horizon and an additional scale, so the wave dynamics depends on both local black-hole structure and global expansion. A substantial body of work has examined QNMs of asymptotically de Sitter geometries across different perturbing fields, parameter ranges, and analytic or numerical methods. A recurring result is the presence of multiple dynamical sectors, including modes continuously connected to black-hole families and modes continuously connected to pure de Sitter space~\cite{RefSdSBranches2024,RefLopezOrtega2012,LopezOrtega:2006qnds,LopezOrtega:2006abs,LopezOrtega:2007dirac}.

Representative milestones include early analyses of Schwarzschild--de Sitter and Reissner--Nordstr\"om--de Sitter spectra, including fundamental and high-overtone behavior for scalar and Dirac and gravitational  perturbations~\cite{Zhidenko:2003wq,Konoplya:2004uk,Molina:2003ff,Jing:2003wq,Stuchlik:2025mjj}. These analyses were then extended to higher-dimensional and brane-localized settings, as well as to stability studies of multidimensional and charged BHs in de Sitter backgrounds~\cite{Kanti:2005xa,Konoplya:2007jv,Konoplya:2008au}.

In modified-gravity sectors, Einstein--Gauss--Bonnet BHs in a de Sitter environment were shown to exhibit qualitatively new mode patterns and instabilities~\cite{Cuyubamba:2016cug,Aragon:2020qdc}. Complementary advances include rigorous results on Kerr--de Sitter mode structure and energy decay, the asymptotic distribution of Kerr--de Sitter resonances, and analyticity/full-subextremal wave-equation analyses in Kerr--de Sitter backgrounds~\cite{Dyatlov:2010hq,Dyatlov:2011asymptotic,Petersen:2021AnalyticityKerrdS,PetersenVasy:2021WaveKdS}, high-accuracy numerical treatments and overdamped modes~\cite{Jansen:2017oag}, stability analyses of QN spectra under perturbations in the presence of a positive cosmological constant~\cite{Sarkar:2023PerturbingQNM}, and strong-cosmic-censorship-motivated analyses in charged de Sitter BHs~\cite{Mo:2018nnu}.

Additional analytic and semi-analytic progress includes near-extremal formulas for Schwarzschild--de Sitter and Kerr--Newman--de Sitter spectra~\cite{CardosoLemos:2003NearExtremalSdS,Churilova:2022NearExtremeKNdS}, classification of asymptotic QN frequencies in higher dimensions~\cite{NatarioSchiappa:2004AsymptoticQNM}, de Sitter-specific quantization/symmetry constructions for QN spectra~\cite{Jafferis:2013dSQNM,Berens:2022LadderQNM}, and recent improvements in nonoscillatory-mode analysis and high-precision spectral methods~\cite{KonoplyaZhidenko:2022Nonosc,KonoplyaZhidenko:2023Bernstein}. Instability analyses for charged and Gauss--Bonnet de Sitter BHs further complete this picture~\cite{KonoplyaZhidenko:2014RNdSInstability,KonoplyaZhidenko:2017EikonalGB}.

Recent work on QN ringing in dark-matter-inspired Weyl gravity found a three-branch structure (Schwarzschild-like, dark-matter, and de Sitter-like), with an exact hypergeometric treatment available in the massless Mannheim--Kazanas limit~\cite{Konoplya:2025WeylQNM}. Related vector-field dark-matter constructions were also studied in~\cite{Heisenberg:2014rta,Fernandes:2025lon}.

Taken together, these results motivate a systematic separation between model-independent de Sitter-wave effects and theory-dependent corrections. We therefore use the exact de Sitter vacuum as a reference background and express the spectrum in terms of the couplings of the underlying gravity theory.

Further developments relevant for asymptotically de Sitter dynamics include early field-propagation analyses in dS BHs~\cite{Molina:2003dc,Dubinsky:2024hmn,Konoplya:2014lha,Konoplya:2003dd}, decay studies for charged fields in Kerr--Newman--de Sitter geometries~\cite{Konoplya:2007zx}, and more recent work on late-time tails, QNM/grey-body correspondences, and comparative spectral behavior across dimensions and asymptotics~\cite{Bolokhov:2024ixe,Malik:2024cgb,Skvortsova:2023zmj}. Recent semi-open-system formulations for Schwarzschild--de Sitter QNMs and related grey-body computations in neighboring models further extend this line of work~\cite{Wu:2025wbp,Dubinsky:2024gwo,Wang:2004bv}.

On the vector-field side, massive Proca perturbations and tails in black-hole backgrounds were established in early studies~\cite{Konoplya:2005hr,Konoplya:2006gq}, while current work emphasizes primary-hair sectors and their phenomenology, including echoes, long-lived modes, optics, and broader Proca-corrected gravity scenarios~\cite{Konoplya:2025uiq,Konoplya:2025bte,Lutfuoglu:2025qkt,Lutfuoglu:2025ldc,Fernandes:2025mic,Fernandes:2026rjs}.

This two-sector picture is particularly clear in the small-black-hole regime,
$
r_h/r_c \ll 1,
$
where $r_h$ and $r_c$ are the event and cosmological horizon radii. In this limit, one branch remains close to asymptotically flat black-hole modes, while the de Sitter-like branch is well approximated by exact pure-de-Sitter frequencies. Consequently, exact modes in the pure de Sitter background provide a useful leading approximation to one branch of nearby black-hole spectra.

The generalized Proca setup considered in~\cite{Heisenberg:2014rta,Fernandes:2025mic,RefProcaGB2026,DeFelice:2016yws,Heisenberg:2016eld,Heisenberg:2017xda,deFelice:2017paw} is suitable for this analysis because it dynamically induces an effective positive cosmological constant, $\Lambda_{\rm eff}>0$, and includes a pure de Sitter vacuum at $M=Q=0$. Its explicit coupling-dependent form is derived in Sec.~III after introducing the composite parameters. Our aim is to present the analytic structure of this de Sitter sector in explicit form. We first review the static branch and its de Sitter asymptotics, then isolate the $M=Q=0$ vacuum, solve the scalar Klein--Gordon equation by hypergeometric reduction, and finally rewrite the exact frequencies directly in terms of the original theory parameters. This also clarifies the transition between purely damped and oscillatory-damped regimes.

The paper is organized as follows. Section~II summarizes the generalized Proca framework and branch conditions. Section~III derives the asymptotically de Sitter structure of the static solutions. Section~IV isolates the pure de Sitter vacuum and its horizon geometry. Section~V presents the full analytic Klein--Gordon derivation and coupling-parameter spectrum. Section~VI discusses physical implications and outlook.

\textbf{Generalized Proca Framework and Branch Conditions.}
Before entering the explicit solution, let us briefly recall the physical motivation for this theory sector. In generalized Proca gravity, the vector field can carry genuine black-hole ``primary hair'', i.e. independent parameters that are not fixed by Gauss-law-type charges at infinity. In the branch studied here, this same structure also allows an effective de Sitter scale to emerge dynamically from the field equations, without inserting a bare cosmological-constant term by hand.

For the purposes of this article, the same integration constants that encode primary-hair information also control the effective asymptotic curvature scale. This gives a direct relation between background solution data and QN observables.

We start from the four-dimensional generalized Proca action~\cite{Heisenberg:2014rta,RefProcaGB2026}
\begin{equation}\nonumber
\begin{split}
S[g,W]=\int \mathrm{d}^4x\,\sqrt{-g}\bigl[
R + c_1 G_{\mu\nu}W^\mu W^\nu + c_2 (W^2)^2 \\
+ c_3 W^2\nabla_\mu W^\mu
\bigr],
\end{split}
\end{equation}
with metric $g_{\mu\nu}$ and vector field $W_\mu$ ($W^2\equiv W_\mu W^\mu$). Here $G_{\mu\nu}\equiv R_{\mu\nu}-\tfrac{1}{2}R g_{\mu\nu}$ is the Einstein tensor, so the $c_1$ term represents a nonminimal Einstein-tensor coupling to the vector field. Variation with respect to $(g_{\mu\nu},W_\mu)$ yields coupled gravitational and vector equations. For static spherical symmetry,
\begin{equation}
\begin{split}
\mathrm{d}s^2&=-h(r)\,\mathrm{d}t^2+\frac{\mathrm{d}r^2}{f(r)}+r^2\mathrm{d}\Omega_2^2,\\
W_\mu \mathrm{d}x^\mu&=w_0(r)\,\mathrm{d}t+w_1(r)\,\mathrm{d}r,
\end{split}
\end{equation}
\begin{equation}
W^2=\frac{w_1^2}{f}-\frac{w_0^2}{h},
\end{equation}
a consistent branch correspond to
\begin{equation}
f(r)=h(r),
\qquad
W^2=\lambda=\text{constant}.
\end{equation}
The constant $\lambda$ is an integration parameter and is not fixed a priori by the Lagrangian couplings.

It is convenient to define
\begin{equation}
\alpha\equiv -\frac{c_1^3}{c_3^2},
\qquad
\beta\equiv \left(1-\frac{8c_1c_2}{3c_3^2}\right)c_1,
\qquad
\lambda=W^2,
\end{equation}
which reorganize the solution in a compact form.

\textbf{Static Solutions and de Sitter Asymptotics.}
In this section we extract the asymptotic physics encoded in the exact static branch. The main goal is to separate which combinations of integration constants and couplings act as effective mass data and which combinations generate the asymptotic de Sitter curvature scale.

The static branch can be written as \cite{RefProcaGB2026,Fernandes:2025mic}
\begin{equation}
\label{eq:masterf2}
\begin{split}
f(r)=1-\frac{2(M-Q)}{r}+\frac{r^2}{2\alpha}
\Bigg(1-\frac{\lambda\beta}{2} \\
\pm\sqrt{1-\lambda\beta\left(1-\frac{c_1\lambda}{4}\right)
+\frac{8\alpha}{r^3}\left[Q+\frac{\lambda(M-Q)}{2\beta}\right]}\Bigg).
\end{split}
\end{equation}

We focus on the minus branch and define~\cite{RefProcaGB2026}
\begin{equation}
A\equiv 1-\frac{\beta\lambda}{2},
\qquad
B\equiv 1-\beta\lambda\left(1-\frac{c_1\lambda}{4}\right),
\end{equation}
where $A$ and $B$ are convenient composite parameters that control, respectively, the effective cosmological term and the reality/regularity of the square-root structure. This rewriting makes the physical parameter constraints more explicit than the original form.
With these definitions, the metric function takes the compact form
\begin{equation}
\label{eq:fminus2}
\begin{split}
f(r)=1-\frac{2(M-Q)}{r}
+\frac{r^2}{2\alpha}\Biggl(A \\
-\sqrt{B+\frac{8\alpha}{r^3}\left[Q+\frac{\lambda(M-Q)}{2\beta}\right]}
\Biggr).
\end{split}
\end{equation}

To make the asymptotics explicit, introduce
\begin{equation}
C\equiv Q+\frac{\lambda(M-Q)}{2\beta},
\end{equation}
and expand for large $r$,
\begin{equation}
\sqrt{B+\frac{8\alpha C}{r^3}}
=\sqrt{B}+\frac{4\alpha C}{\sqrt{B}\,r^3}+\mathcal{O}(r^{-6}).
\end{equation}
Substituting into \eqref{eq:fminus2} gives
\begin{equation}\nonumber
\begin{split}
f(r)=1-\frac{2}{r}\left[(M-Q)+\frac{C}{\sqrt{B}}\right]
+\frac{r^2}{2\alpha}(A-\sqrt{B}) \\
+\mathcal{O}(r^{-4}).
\end{split}
\end{equation}
Hence
\begin{equation}
f(r)=1-\frac{2M_{\rm eff}}{r}-\frac{\Lambda_{\rm eff}}{3}r^2+\mathcal{O}(r^{-4}),
\end{equation}
%with
\begin{equation}
\begin{split}
M_{\rm eff}&=(M-Q)+\frac{1}{\sqrt{B}}\left[Q+\frac{\lambda(M-Q)}{2\beta}\right],\\
\Lambda_{\rm eff}&=\frac{3}{2\alpha}(\sqrt{B}-A).
\end{split}
\end{equation}
The asymptotically de Sitter sbranch corresponds to
\begin{equation}
\label{eq:dScond2}
B\ge 0,
\qquad
\Lambda_{\rm eff}>0.
\end{equation}

\textbf{Pure de Sitter Vacuum and Horizon Structure.}
We now isolate the background obtained by removing black-hole integration data. This step is physically important because it identifies the exact geometry that governs the de Sitter-like QNM sector in the small-black-hole limit.

Setting
$M=0, Q=0,$
Eq.~\eqref{eq:fminus2} reduces exactly to
\begin{equation}
\label{eq:fM0Q0}
f_0(r)=1+\frac{r^2}{2\alpha}(A-\sqrt{B})
=1-\frac{\Lambda_{\rm eff}}{3}r^2.
\end{equation}
In the domain \eqref{eq:dScond2}, this is the static patch of pure de Sitter space,
\begin{equation}\nonumber
H\equiv \sqrt{\frac{\Lambda_{\rm eff}}{3}},
\qquad
f_0(r)=1-H^2r^2,
\qquad
r_c=H^{-1}.
\end{equation}

The corresponding tortoise coordinate is
\begin{equation}
r_*\equiv \int \frac{\mathrm{d}r}{1-H^2r^2}
=\frac{1}{2H}\ln\left(\frac{1+Hr}{1-Hr}\right),
\end{equation}
which diverges logarithmically as $r\to r_c^-$. This makes wave boundary conditions at the cosmological horizon directly analogous to standard one-dimensional scattering problems. Since the $1/r$ mass term is absent, this vacuum provides a useful analytic baseline for the de Sitter-like QN sector of nearby black-hole solutions.

Operationally, this means that the pure-vacuum problem supplies the zeroth-order spectrum, while nonzero $(M,Q)$ can be treated as perturbations in subsequent analytic or numerical studies.

\textbf{Scalar Perturbations: Detailed Derivation of the Spectrum.}
Having identified the exact vacuum, we now derive the scalar spectrum analytically. In this derivation, we follow Ref.~\cite{RefLopezOrtega2012}: the strategy is to reduce the wave equation to a hypergeometric problem and then impose boundary conditions at the origin and cosmological horizon.
This derivation is useful not only because it yields closed-form frequencies, but also because each step shows the origin of quantization conditions: regularity at the origin and one-way flux at the cosmological horizon.
From a methodological viewpoint, this analytic solvability is non-generic in black-hole perturbation theory: most physically relevant geometries require numerical eigenvalue extraction.
We consider a scalar field with mass $\mu$ obeying
$
(\Box-\mu^2)\Phi=0,
$
in the background \eqref{eq:fM0Q0}.
%\textbf{Radial Equation and Effective Potential.}
The covariant wave equation can be converted into a one-dimensional scattering problem. Following Ref.~\cite{RefLopezOrtega2012}, the effective potential form is then used to identify the terms governing trapping, transmission, and late-time damping. Using
\begin{equation}
\Phi=e^{-i\omega t}Y_{\ell m}(\theta,\varphi)\frac{U(r)}{r},
\end{equation}
a straightforward separation gives (see, for example, \cite{Kamran:1984mb,Konoplya:2018arm})
\begin{equation}
\begin{split}
\frac{1}{r^2}\frac{\mathrm{d}}{\mathrm{d}r}\left(r^2f_0\frac{\mathrm{d}}{\mathrm{d}r}\frac{U}{r}\right)
+\Biggl(\frac{\omega^2}{f_0}-\frac{\ell(\ell+1)}{r^2} \\
-\mu^2\Biggr)\frac{U}{r}=0.
\end{split}
\end{equation}
After rewriting in tortoise coordinate,
\begin{equation}
\frac{\mathrm{d}r_*}{\mathrm{d}r}=\frac{1}{f_0(r)}=\frac{1}{1-H^2r^2},
\end{equation}
one obtains Schr\"odinger form,
\begin{equation}
\frac{\mathrm{d}^2U}{\mathrm{d}r_*^2}+\left(\omega^2-V_{\rm KG}(r)\right)U=0,
\end{equation}
%with
\begin{equation}
V_{\rm KG}(r)=f_0(r)\left[\frac{\ell(\ell+1)}{r^2}+\mu^2-2H^2\right].
\end{equation}

\begin{figure}
%\centering
\includegraphics[width=0.47\textwidth]{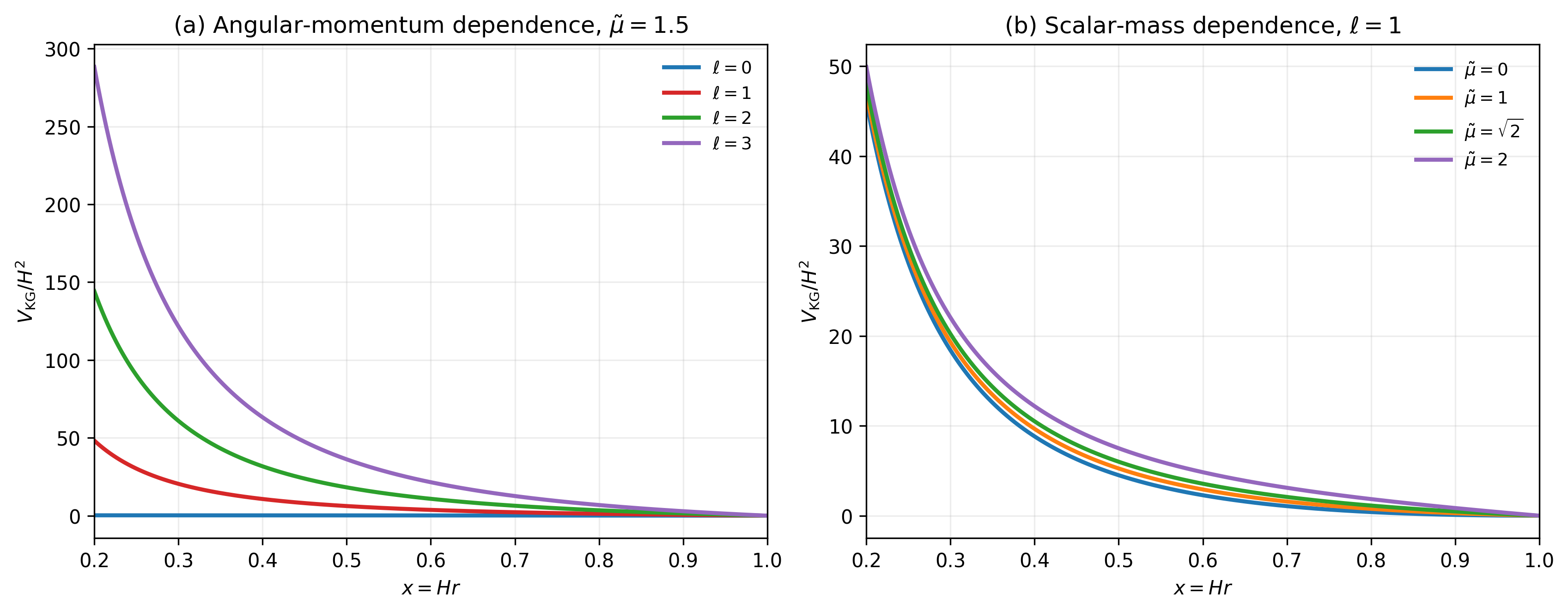}
\caption{Dimensionless Klein--Gordon effective potential $V_{\rm KG}/H^2$. Left: angular-momentum dependence for fixed $\tilde\mu=\mu/H=1.5$. Right: scalar-mass dependence for fixed $\ell=1$. The radial range starts at $x=0.2$ to display the potential away from the centrifugal divergence at the origin for $\ell>0$; all curves vanish at the cosmological horizon $x=1$.}
\label{fig:eq25-field-scans}
\end{figure}

We introduce dimensionless variables
\begin{equation}
x\equiv Hr\in(0,1),
\qquad
\tilde\omega\equiv \frac{\omega}{H},
\qquad
\tilde\mu\equiv \frac{\mu}{H},
\end{equation}
and obtain
\begin{equation}
\label{eq:radx}
\begin{split}
(1-x^2)U''-2xU'+\Biggl[\frac{\tilde\omega^2}{1-x^2}
-\frac{\ell(\ell+1)}{x^2} \\
+(2-\tilde\mu^2)\Biggr]U=0.
\end{split}
\end{equation}
\textbf{Hypergeometric Reduction and Boundary Conditions.}
As in Ref.~\cite{RefLopezOrtega2012}, the next step is to map the radial equation to a standard special-function form, so that quantization conditions can be imposed algebraically rather than numerically. The boundary conditions correspond to regular behavior at the center and outgoing propagation at the cosmological horizon.
Setting $y=x^2$ gives
\begin{equation}
\begin{split}
y(1-y)U_{yy}+\Biggl(\frac{1}{2}-\frac{3}{2}y\Biggr)U_y \\
+\left[\frac{\tilde\omega^2}{4(1-y)}-\frac{\ell(\ell+1)}{4y}+\frac{2-\tilde\mu^2}{4}\right]U=0.
\end{split}
\end{equation}
Now use
\begin{equation}
U=(1-y)^E y^C U_2(y),
\end{equation}
with indicial equations
\begin{equation}
E^2+\frac{\tilde\omega^2}{4}=0,
\qquad
C^2-\frac{C}{2}-\frac{\ell(\ell+1)}{4}=0.
\end{equation}

At the origin ($x\to0$), regularity selects
\begin{equation}
C=\frac{\ell+1}{2}.
\end{equation}
At the cosmological horizon ($x\to1$), the local behavior is
\begin{equation}
U\sim(1-x^2)^{-i\tilde\omega/2}\propto e^{+i\omega r_*},
\end{equation}
which corresponds to outgoing waves; thus
\begin{equation}
E=-\frac{i\tilde\omega}{2}.
\end{equation}
The reduced function $U_2$ is hypergeometric, and the radial mode is
\begin{equation}
U(x)=x^{\ell+1}(1-x^2)^{-i\tilde\omega/2}\,{}_2F_1(a,b;c;x^2),
\end{equation}
with
\begin{equation}
\begin{aligned}
a&=\frac{\ell}{2}+\frac{3}{4}-\frac{1}{2}\nu-\frac{i\tilde\omega}{2},\\
b&=\frac{\ell}{2}+\frac{3}{4}+\frac{1}{2}\nu-\frac{i\tilde\omega}{2},\\
c&=\ell+\frac{3}{2},
\end{aligned}
\qquad
\nu\equiv\sqrt{\frac{9}{4}-\tilde\mu^2}.
\end{equation}

\textbf{Exact Spectrum and Physical Regimes.}
With the hypergeometric structure fixed, the spectrum follows from polynomial truncation, following Ref.~\cite{RefLopezOrtega2012}. This yields exact expressions and allows a direct classification of modes into purely damped and oscillatory-damped regimes.

Polynomial truncation,
\begin{equation}
a=-n\quad\text{or}\quad b=-n,
\qquad n=0,1,2,\dots,
\end{equation}
leads to the exact frequencies
\begin{equation}
\label{eq:omegafinal}
\begin{split}
\tilde\omega_{n\ell}^{(\pm)}&=-i\left(2n+\ell+\frac{3}{2}\pm\nu\right),\\
\omega_{n\ell}^{(\pm)}&=-iH\left(2n+\ell+\frac{3}{2}\pm\nu\right).
\end{split}
\end{equation}

For $\mu=0$ ($\nu=3/2$),
\begin{equation}
\omega_{n\ell}^{(1)}=-iH(2n+\ell),
\qquad
\omega_{n\ell}^{(2)}=-iH(2n+\ell+3).
\end{equation}

The QNMs follows directly from $\nu$:
\begin{equation}
0\le\mu^2<\frac{9H^2}{4}
\quad\Rightarrow\quad
\nu\in\mathbb{R}
\ \text{and}\ \omega\ \text{purely imaginary},
\end{equation}
while for heavier fields,
\begin{equation}
\mu^2>\frac{9H^2}{4},
\qquad
\nu=i\eta,\ \eta>0,
\end{equation}
one has oscillatory damping,
\begin{equation}
\omega_{n\ell}^{(\pm)}=\pm H\eta-iH\left(2n+\ell+\frac{3}{2}\right).
\end{equation}
The exceptional point $(D,\ell,\tilde\mu^2)=(4,0,2)$ has $V_{\rm KG}\equiv0$ and requires separate treatment~\cite{RefLopezOrtega2012}.

\textbf{Spectrum in Terms of Original Theory Parameters.}
Finally, we translate the frequency formulas back to the original Proca-theory parameters. This step is essential for interpretation, because it connects observable damping and oscillation patterns to the couplings that define the gravitational background. 
Using
\begin{equation}
\begin{split}
H^2&=\frac{\Lambda_{\rm eff}}{3}=\frac{\sqrt{B}-A}{2\alpha},\\
A&=1-\frac{\beta\lambda}{2},
\qquad B=1-\beta\lambda\left(1-\frac{c_1\lambda}{4}\right),
\end{split}
\end{equation}
we obtain\\
\begin{equation}\nonumber
\omega_{n\ell}^{(\pm)}=-i\sqrt{\frac{\sqrt{B}-A}{2\alpha}}
\left(2n+\ell+\frac{3}{2}\pm\sqrt{\frac{9}{4}-\frac{2\alpha\mu^2}{\sqrt{B}-A}}\right),
\end{equation}
or explicitly in terms of $(\alpha,\beta,c_1,\lambda)$, 
\vspace{6mm}
\begin{widetext}
\begin{equation}\nonumber
\omega_{n\ell}^{(\pm)}=-i\sqrt{\frac{\sqrt{1-\beta\lambda\left(1-\frac{c_1\lambda}{4}\right)}-\left(1-\frac{\beta\lambda}{2}\right)}{2\alpha}}\times\left(2n+\ell+\frac{3}{2}\pm\sqrt{\frac{9}{4}-\frac{2\alpha\mu^2}{\sqrt{1-\beta\lambda\left(1-\frac{c_1\lambda}{4}\right)}-\left(1-\frac{\beta\lambda}{2}\right)}}\right).
\end{equation}
\end{widetext}

\vspace{6mm}
For the massless field, we have
\begin{equation}
\begin{split}
\omega_{n\ell}^{(1)}&=-i\sqrt{\frac{\sqrt{B}-A}{2\alpha}}(2n+\ell),\\
\omega_{n\ell}^{(2)}&=-i\sqrt{\frac{\sqrt{B}-A}{2\alpha}}(2n+\ell+3).
\end{split}
\end{equation}

\textbf{Discussion and Conclusions.}
QNMs can usually be determined only numerically because of the complexity of the master wave equation. Although a few exact solutions are known, exact QN spectra are unavailable not only for the simplest four-dimensional Schwarzschild black hole, but even for lower-dimensional BHs, such as BTZ-like solutions in quantum-corrected theories \cite{Konoplya:2020ibi,Skvortsova:2023zca,Skvortsova:2023zmj}. Therefore, any new analytic solution for QN spectra is of considerable theoretical interest \cite{Bolokhov:2025uxz,Konoplya:2023moy}.

After the first version of this work appeared on arXiv, further investigations of this framework were carried out in \cite{Skvortsova:2026idf,Lutfuoglu:2026pgn,Bolokhov:2026dfg}, where the QN spectra of BHs in the same theory were studied. It was demonstrated that, as predicted in the present work, the de Sitter branch of QNMs for BHs whose event-horizon radius is much smaller than the de Sitter radius agrees remarkably well with the analytic expressions derived here. The principal tool employed there was the WKB method, which has also been used in numerous recent studies \cite{Media:2025udn,Devi:2026tsr,Singh:2025tvk,Singh:2025fuv,Konoplya:2024hfg,Konoplya:2019ppy,Bolokhov:2023bwm,Bolokhov:2025fto,Bolokhov:2025lnt,Konoplya:2025hgp,Konoplya:2023ppx,Bolokhov:2025egl,Bolokhov:2023ruj,Konoplya:2024kih,Skvortsova:2024atk,Skvortsova:2024eqi,Skvortsova:2024wly,Lutfuoglu:2026zxj,Lutfuoglu:2026rqe,Tan:2026itp}. However, the WKB method cannot reproduce the analytic QNMs derived here because the effective potential does not have a WKB-canonical single peak shape (see fig. \ref{fig:eq25-field-scans}). Consequently, After all, the purely de Sitter modes cannot be recovered within the WKB approximation, as shown in \cite{Konoplya:2022gjp}, nor by conceptually similar approaches such as the Pöschl--Teller potential fit. The pseudospectral method, however, could be used here \cite{Fortuna:2020obg,Konoplya:2024lch}, because it does not depend on the potential's shape.

The de Sitter sector of the generalized Proca branch provides an analytic setting in which effective-cosmological-constant physics and black-hole perturbations can be related explicitly. At the geometric level, the parameters $(M,Q)$ determine the interior branch structure, while $(\alpha,\beta,c_1,\lambda)$ determine the asymptotic de Sitter scale, with $c_1$ entering through the Einstein-tensor coupling $c_1 G_{\mu\nu}W^\mu W^\nu$. At the level of wave dynamics, the $M=Q=0$ vacuum captures the leading de Sitter-like spectral behavior of nearby small-black-hole configurations.

The explicit mode formulas obtained here show how couplings and scalar mass control both damping rates and the onset of oscillatory behavior. This provides a basis for perturbative analyses around small but nonzero $(M,Q)$, where one can track how pure de Sitter modes are modified in the full asymptotically de Sitter black-hole spectrum.

A complementary perspective concerns strong cosmic censorship (SCC) in asymptotically de Sitter BHs, where the competition between exterior decay and Cauchy-horizon blueshift is controlled by the dominant QN frequency. A standard diagnostic is $\beta_{\rm SCC}\equiv-\mathrm{Im}(\omega_0)/\kappa_-$, with potential Christodoulou-SCC violation when $\beta_{\rm SCC}>1/2$ for sufficiently smooth data. Foundational analyses of Reissner--Nordstr\"om--de Sitter perturbations established this criterion and showed near-extremal parameter regions where SCC can be challenged, while rough-data formulations significantly tighten the verdict in favor of SCC~\cite{Cardoso:2017soq,Dias:2018etb,Mo:2018nnu}. In a broader survey across asymptotically de Sitter geometries, Konoplya and Zhidenko found that for relatively small BHs the dominant modes satisfy both $-\mathrm{Im}(\omega)\le\kappa_+/2$ and $-\mathrm{Im}(\omega)\le\kappa_-/2$, i.e. the Hod-type relaxation bound and the SCC-motivated bound simultaneously~\cite{KonoplyaZhidenko:2022SCC}. In this light, the exact de Sitter spectrum derived here [Eq.~\eqref{eq:omegafinal}] provides an analytic baseline for the de Sitter-like branch entering those SCC diagnostics once small but finite $(M,Q)$ deformations are restored.  We conclude that BHs with an event-horizon radius much smaller than the de Sitter radius necessarily satisfy both the Hod and strong cosmic censorship (SCC) bounds on the damping rate. Indeed, the damping rate scales inversely with the de Sitter radius and thus becomes arbitrarily small in the limit of a large de Sitter radius.

\textbf{Acknowledgements.} 
I am grateful to Prof. Lavinia Heisenberg for drawing my attention to an important work on generalized Proca theory.

\bibliographystyle{apsrev4-1}
\bibliography{referencesProca}

\end{document}